\documentclass[proof]{pasj00}

\begin{document}
\SetRunningHead{Hirota et al.}{VERA astrometry of L1448C}
\Received{2010/06/01}
\Accepted{2010/XX/XX}

\title{Astrometry of H$_{2}$O Masers in Nearby Star-Forming Regions with VERA 
--- IV. L1448C}

\author{
Tomoya \textsc{Hirota},\altaffilmark{1,2}
Mareki \textsc{Honma},\altaffilmark{1,2}
Hiroshi \textsc{Imai},\altaffilmark{3} 
Kazuyoshi \textsc{Sunada},\altaffilmark{2,4} \\
YuJi \textsc{Ueno},\altaffilmark{4} 
Hideyuki \textsc{Kobayashi},\altaffilmark{1,5}
Noriyuki \textsc{Kawaguchi},\altaffilmark{2,4} \\
}
\altaffiltext{1}{Mizusawa VLBI Observatory, National Astronomical Observatory of Japan, \\
  2-21-1 Osawa, Mitaka, Tokyo 181-8588}
\altaffiltext{2}{Department of Astronomical Sciences, Graduate University for Advanced Studies, \\
  2-21-1 Osawa, Mitaka, Tokyo 181-8588}
\altaffiltext{3}{Graduate School of Science and Engineering, Kagoshima University, \\
  1-21-35 Korimoto, Kagoshima, Kagoshima 890-0065}
\altaffiltext{4}{Mizusawa VLBI Observatory, National Astronomical Observatory of Japan, \\
  2-12 Hoshi-ga-oka, Mizusawa-ku, Oshu-shi, Iwate 023-0861}
\altaffiltext{5}{Department of Astronomy, Graduate School of Science, The University of Tokyo, \\
  7-3-1 Hongo, Bunkyo-ku, Tokyo 113-0033}
\email{tomoya.hirota@nao.ac.jp}

\maketitle

\begin{abstract}
We have carried out multi-epoch VLBI observations with VERA (VLBI Exploration of 
Radio Astrometry) of the 22~GHz H$_{2}$O masers associated with a Class 0 protostar 
L1448C in the Perseus molecular cloud. 
The maser features trace the base of collimated bipolar jet driven by one of the infrared 
counter parts of L1448C named as L1448C(N) or L1448-mm A. We detected possible evidences for 
apparent acceleration and precession of the jet according to the three-dimensional velocity structure. 
Based on the phase-referencing VLBI astrometry, 
we have successfully detected an annual parallax 
of the H$_{2}$O maser in L1448C to be 4.31$\pm$0.33~milliarcseconds (mas) 
which corresponds to a distance of 232$\pm$18~pc from the Sun. 
The present result is in good agreement with that of another H$_{2}$O maser source 
NGC~1333 SVS13A in the Perseus molecular cloud, 235~pc. 
It is also consistent with the photometric distance, 220~pc. 
Thus, the distance to the western part of the Perseus molecular cloud complex would be 
constrained to be about 235~pc rather than the larger value, 300~pc, previously reported. 
\end{abstract}

\section{Introduction}

In order to understand formation processes of stars, 
it is necessary to obtain accurate physical and dynamical properties of 
newly born young stellar objects (YSOs) 
such as size, mass, and luminosity. For this purpose, 
an accurate distance to a YSO in a star-forming region 
is the most fundamental parameter for quantitative discussion. 
In the last decade, great efforts have been made to establish the highest 
accuracy astrometric observations with the very long baseline interferometry (VLBI). 
This, VLBI astrometry, enables to measure annual parallaxes 
for bright radio sources, i.e. maser sources and non-thermal radio sources, associated 
with YSOs at a typical accuracy of sub-milliarcseconds (mas). 
It can achieve much better accuracy by 2-3 orders of magnitude than that of 
the optical astrometry satellite Hipparcos \citep{perryman1997}. 
As a result, the distances to several well studied star-forming regions such as 
Taurus \citep{loinard2005, loinard2007, torres2007, torres2009}, 
Ophiuchus \citep{imai2007, loinard2008}, 
Orion \citep{hirota2007, sandstrom2007, menten2007, kim2008}, 
Perseus \citep{hirota2008a}, and Cepheus\citep{hirota2008b, moscadelli2009} 
regions were refined with better than a few percent uncertainties. 

Among them, we have been carrying out a VLBI astrometry project 
"Measurements of annual parallaxes of nearby molecular clouds" with 
VERA (VLBI Exploration of Radio Astrometry). VERA is a Japanese VLBI network 
operated by National Astronomical Observatory of Japan (NAOJ) and Kagoshima University. 
VERA is designed to dedicate for astrometric observations aimed at 
revealing three-dimensional structure of the Galaxy (e.g. \cite{honma2007}). 
Our project mainly focuses on the accurate distance measurements 
of star-forming regions within 1~kpc from the Sun \citep{dame1987, evans2003}. 
Part of the results have been reported in a series of papers 
\citep{hirota2007, imai2007, hirota2008a, hirota2008b, kim2008}. 

\begin{figure*}[hbt]
  \begin{center}
    \FigureFile(80mm,80mm){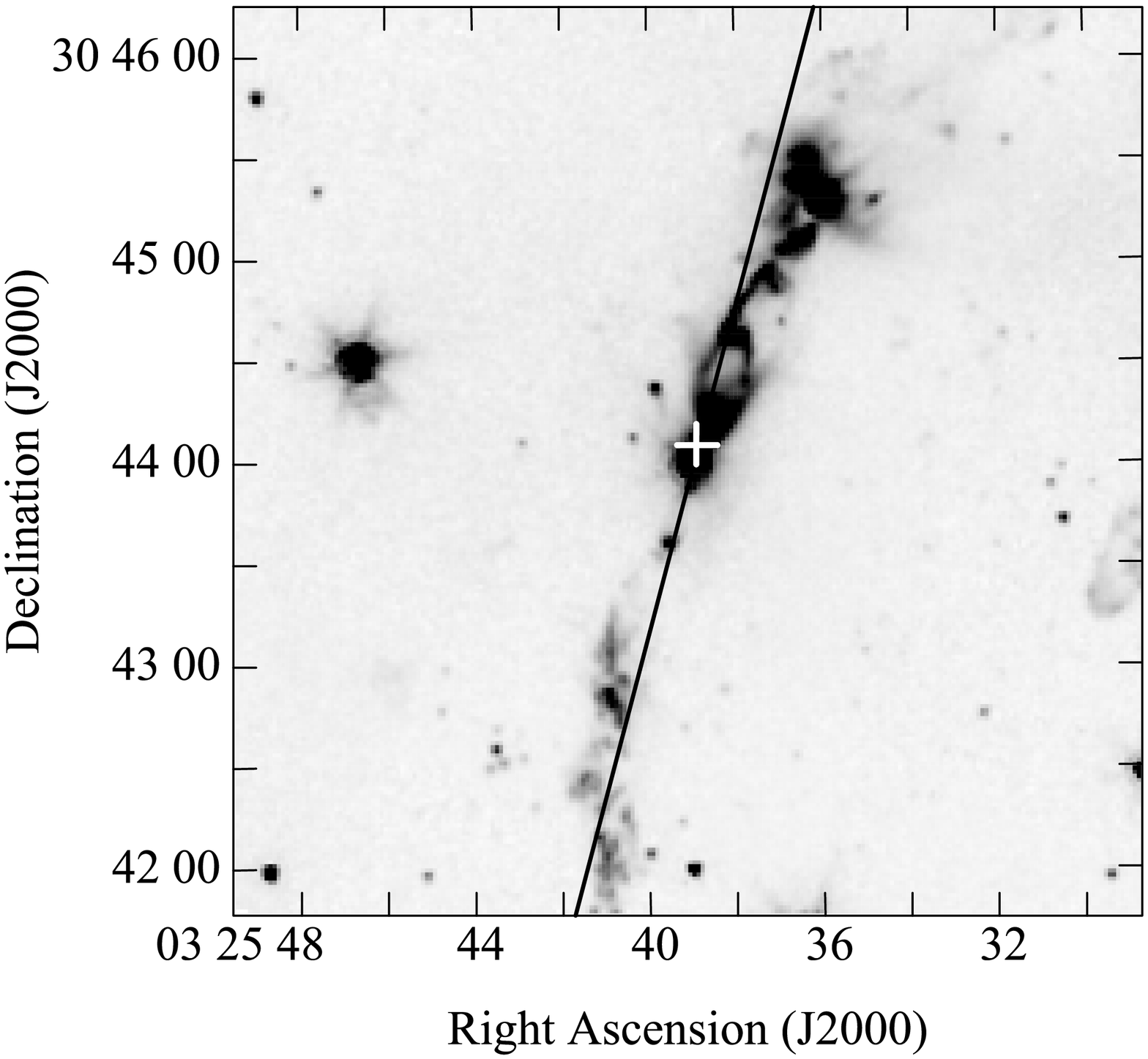}
    \FigureFile(80mm,80mm){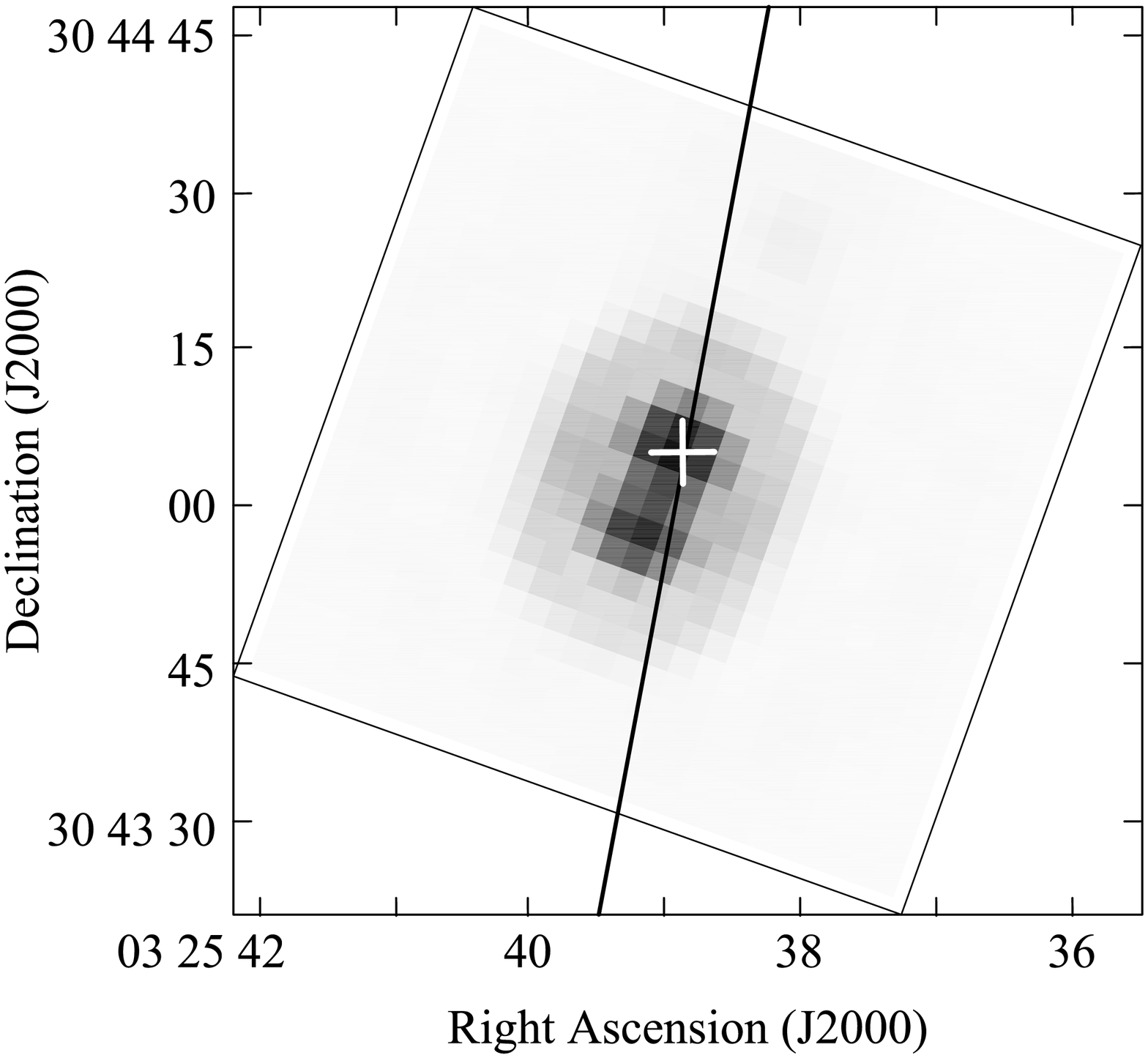}
  \caption{Spitzer IRAC channel 2 image (left) and MIPS channel 1 image (right) 
   of L1448C taken from the Spitzer Legacy 
   Program "From Molecular Cores to Planet Forming Disks" 
   \citep{evans2003}. 
   A white cross represents the reference position adopted in the present study. 
   A black solid line represents the average position angle of the H$_{2}$O maser 
   features in L1448C.   }
  \label{fig-spitzer}
  \end{center}
\end{figure*}

Here we present a new result of our VLBI astrometry of the H$_{2}$O masers 
associated with L1448C (or L1448-mm) in an optical dark cloud Lynds 1448 \citep{lynds1962}. 
L1448 is located at 1~degree southwest of another star-forming region NGC~1333 
\citep{hirota2008a} in the Perseus molecular cloud. 
L1448C was first detected at the millimeter \citep{bachiller1991} and 
centimeter \citep{curiel1990} wavelengths, and was later classified as 
a deeply embedded Class 0 protostar based on its spectral energy 
distribution (SED) from the infrared to millimeter wavelengths \citep{barsony1998}. 
Recent observations with the Spitzer Space Telescope revealed that 
L1448C consists of two infrared counterpart named as 
L1448C(N) and L1448C(S) \citep{jorgensen2006} or 
L1448-mm A and B \citep{tobin2007} 
which are separated by 8\arcsec \ as shown in Figure \ref{fig-spitzer}. 
These infrared sources have been observed more recently with the 
higher resolution interferometers at 
the centimeter to submillimeter wavelengths 
\citep{reipurth2002, hirano2010}. 
L1448C is known to drive large-scale extremely high-velocity 
molecular outflow traced by 
the millimeter rotational lines of the CO and SiO molecules 
\citep{bachiller1990, bachiller1995, girart2001, hirano2010} 
and the vibrationally excited 
H$_{2}$ line in the near infrared wavelength \citep{bally1993}. 
Thus, L1448C is one of the ideal laboratories to study in detail about the 
mass-loss processes through the collimated jet and molecular outflow, as well as 
the nature of the protostar in very young evolutionary phase, and hence, 
the accurate distance measurement is essential for this purpose. 
The H$_{2}$O masers were detected with single-dish telescopes \citep{claussen1996} 
and the Very Large Array (VLA) \citep{chernin1995}. 
However, this is the first time to observe the H$_{2}$O masers associated with L1448C 
with VLBI. The highest resolution observations with VERA yield 
the proper motions of the masers along with the annual parallax of L1448C. 

\section{Observations and Data Analyses}

Observations of the H$_{2}$O maser line ($6_{1 6}$-$5_{2 3}$, 22235.080 MHz) 
associated with L1448C were conducted with VERA from November 2007 to February 2009. 
All 4 stations of VERA (see Fig.1 of \cite{petrov2007}) took part in all observing sessions, 
providing a maximum baseline length of 2270~km. 

Observations were made in the dual beam mode; the H$_{2}$O masers associated 
with L1448C and an extragalactic radio source J0319$+$3101 \citep{petrov2006}, 
with a separation angle of $1.37^{\circ}$, were observed simultaneously. 
The instrumental phase difference between the two beams was measured 
continuously during the observations by injecting 
artificial noise sources into both beams at each station \citep{honma2003, honma2008a}. 

Left-handed circular polarization was received and sampled with 2-bit 
quantization and filtered using the VERA digital filter unit (\cite{iguchi2005}). 
The data were recorded onto magnetic tapes at a rate of 1024~Mbps, 
providing a total bandwidth of 256~MHz in which one IF channel and the rest 
of 15 IF channels with a 16~MHz bandwidth each were assigned to 
L1448C and J0319+3101, respectively. 
A bright extragalactic radio source, 3C84, was observed every 80 minutes 
as a delay and bandpass calibrator. 
Amplitude calibrations were done through the chopper-wheel method (\cite{ulich1976}). 
Correlation processing was carried out on the Mitaka FX correlator 
(\cite{chikada1991}) located at the NAOJ Mitaka campus. 
For the H$_{2}$O maser line, the spectral resolution was set to be 15.625~kHz, 
corresponding to the velocity resolution of 0.21~km~s$^{-1}$. 

Data reduction was performed using 
the National Radio Astronomy Observatory (NRAO) Astronomical Image Processing System (AIPS). 
We first applied the results of the dual-beam phase calibration as mentioned above 
and the correction for the approximate delay model adopted in the correlation 
processing. Details in this procedure are described in previous papers 
\citep{honma2003, honma2007, honma2008a, honma2008b}. 
The reference position of L1448C was set to be the geometric center of 
two maser features as discussed later, 
RA(J2000)=03h25m38.87840s and Dec(J2000)=+30$^{\circ}$44'05".2516, 
in the recalculation of the delay tracking model. 
Next, we calibrated the instrumental delays and phase offsets 
among all of the IF channels by the AIPS task FRING on 3C84. 
Finally, we calibrated residual phases by the AIPS task FRING on J0319+3101. 
The solutions were applied to the target source L1448C. 
The reference source J0319+3101 has a flux density of only 
30~mJy~beam$^{-1}$ at 22~GHz and hence, it 
was only marginally detected with the signal-to-noise ratio of about 5 
during most of the observing sessions. Therefore, we also 
calibrated the residual phases by the AIPS task FRING on the intense 
spectral feature of the H$_{2}$O maser in L1448C instead of J0319+3101. 
In this case, the solutions were also applied to J0319+3101. 
These two results will be compared in the later section. 

Synthesis imaging and deconvolution (CLEAN) were performed using 
the AIPS task IMAGR. The uniform weighted synthesized beam size 
(FWMH) was typically 1.2~mas$\times$0.8~mas with a position angle of -50~degrees. 
The peak positions and flux densities 
of masers were derived by fitting elliptical Gaussian brightness distributions 
to each spectral channel map using the AIPS task SAD. The formal uncertainties in 
the maser positions given by SAD were better than 0.1~mas. 
The rms noise levels in the self-calibrated images are 1-2~mJy~beam$^{-1}$ for 
J0319+3101 and 50-150~mJy~beam$^{-1}$ for the channel maps of 
the L1448C masers with the net integration time of about 3~hours. 

\section{Results}
\subsection{Structure of the H$_{2}$O maser features}

\begin{figure}[hbt]
  \begin{center}
    \FigureFile(80mm,80mm){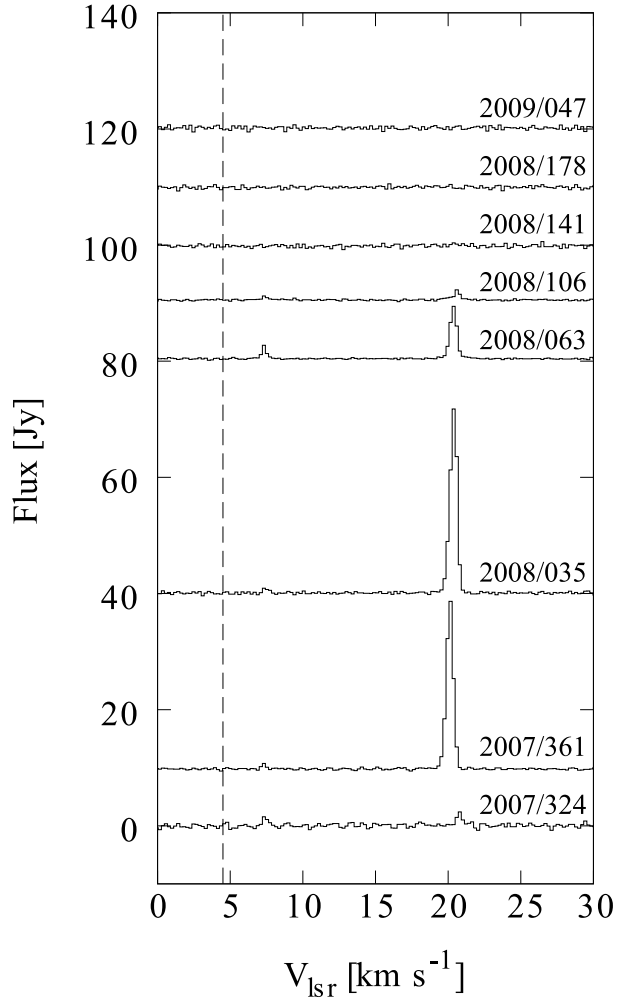}
  \caption{Scalar-averaged cross power spectra of the H$_{2}$O maser lines associated with 
  L1448C. A dashed line represents the systemic velocity of 4.5~km~s$^{-1}$ 
  \citep{bachiller1990}. 
  The spectra are the averages of those observed at all the VERA stations. }
  \label{fig-spectra}
  \end{center}
\end{figure}

Figure \ref{fig-spectra} shows the spectra of the H$_{2}$O masers associated with L1448C. 
We detected the masers at five observing sessions, 
2007/324, 2007/361, 2008/035, 2008/063, and 2008/106 (denoted by year/day of the year). 
For later sessions, 
2008/141, 2008/178, and 2009/047, we could not detect the masers due to the time variation
which is characteristic for the H$_{2}$O masers associated with low-mass protostars 
\citep{claussen1996}. 
The most intense feature was always at the local standard of rest (LSR) velocity of about 20~km~s$^{-1}$. 
The second weak feature was detected at the LSR velocity of 7~km$^{-1}$ from 
2007/324 to 2008/106, although they are marginally seen in Figure \ref{fig-spectra}. 
Both features are red-shifted with respect to the systemic velocity of 
the molecular cloud L1448, 4.5~km~s$^{-1}$ \citep{bachiller1990}. 
The velocities of the masers are within that of the molecular outflow ranging up to 
$\pm$70~km~s$^{-1}$ with respect to the systemic velocity \citep{bachiller1990, bachiller1995}. 
These two features had been identified by \citet{chernin1995} 
whereas we could not detect other velocity components detected by \citet{chernin1995} and 
\citet{claussen1996}. 
We also detected another faint maser feature in 2009/047 at the LSR velocity of 
-24~km~s$^{-1}$. Because this feature could not be detected in the VLBI imaging, 
we will exclude this feature in the following discussion. 

\begin{figure*}[htb]
  \begin{center}
    \FigureFile(170mm,170mm){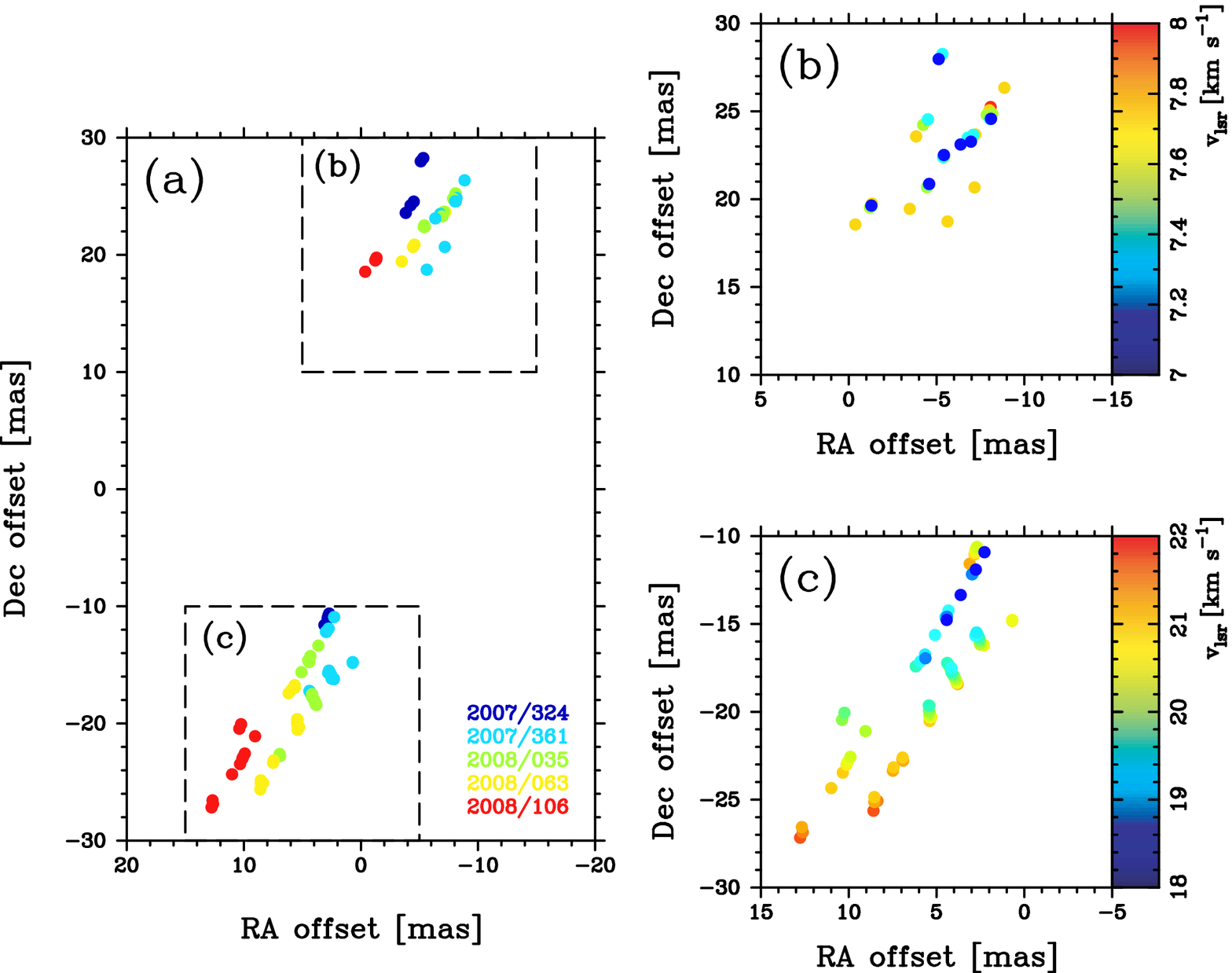}
  \caption{Distribution of H$_{2}$O maser spots associated with L1448C. 
   (a) Overall distribution of the maser spots. Each color represents the observed epoch. 
   (b) Close-up of the northern maser feature as indicated by the dashed rectangle in panel (a). 
   Color-coded symbols represent the positions 
   and LSR velocities of the maser spots for all of the observed epochs. 
   (c) Same as (b) but for the southern maser feature. 
   The reference position, 
   $\alpha(J2000)=$03h25m38.87840s, $\delta(J2000)=+30$d44\arcmin05\arcsec.2516, 
   is determined with respect to the position of J0319+3101. }
  \label{fig-allspot}
  \end{center}
\end{figure*}

The distribution of the maser spots are shown in Figure \ref{fig-allspot}. 
Hereafter we define a ``spot" as emission occurring in a single velocity channel and 
a ``feature" as a group of spots. We define two features; one is located at 
northwest of the reference position while another at southeast. 

In order to obtain absolute positions of the maser spots, 
we at first made synthesis imaging for each epoch by employing the result of 
the phase-calibration on the reference maser spot at the LSR velocity of 20.6~km~s$^{-1}$. 
This is because we could obtain higher sensitivity images compared with those 
obtained by phase-referencing to J0319+3101 or other maser spots at different velocities. 
In this case, positions of the maser spots are 
measured with respect to the reference spot. Next, we transferred the phase-calibration 
results to the J0319+3101 data. 
The absolute position of J0319+3101 has been determined 
with an uncertainty of 1.63 and 1.68~mas in right ascension and declination, respectively 
\citep{petrov2006}, so that we could convert the position offsets of the maser spots 
with respect to J0319+3101 into absolute positions. 

As a result, we can easily locate the absolute position of the maser source to be one of the 
Spitzer sources, L1448C(N)/L1448-mm A 
\citep{jorgensen2006, tobin2007} as shown in Figure \ref{fig-spitzer}. 
The position also agrees well with that of the radio continuum source 
\citep{reipurth2002} and the submillimeter continuum source \citep{hirano2010}. 
Note that the positions of the continuum source obtained by the 
interferometer observations \citep{reipurth2002, hirano2010} tend to be 
shifted toward 0.12-0.17\arcsec \ northwest of our reference position. 
Although the astrometric accuracies seem to be insufficient with the beam sizes of 
0.3\arcsec-0.7\arcsec \ achieved by these observations, 
the powering source of the masers could be located at the northwest of the 
maser images shown in Figure \ref{fig-allspot}. 

The spatial distribution of the H$_{2}$O masers linearly aligned with 
a collimated jet-like structure seems to be analogous to the large scale 
molecular outflow, although the size of the maser structure is much smaller 
by 3-4 orders of magnitude. 
The width of the H$_{2}$O maser jet is less than 5~mas
while its length is 60~mas as can be seen in Figure \ref{fig-allspot}. 
This scale is significantly smaller than that found by \citet{chernin1995}, 280~mas. 
The estimated size gives the lower limit for only part of the 
(red-shifted) jet because we could not detect the blue-shifted H$_{2}$O maser 
jet in the present study. Details of the spatial and velocity structure of 
the maser features will be discussed in the later section. 

\begin{table*}[htb]
\begin{center}
\caption{Position offsets of the reference maser spots ($v_{lsr}$=20.6~km~s$^{-1}$) 
derived from the phase-referencing methods}
\label{tab-prcompare1}
\begin{tabular}{cccc}
\hline
\hline
              &  reference: L1448C$^{a}$ & reference: J0319$^{b}$ & difference \\
epoch      &  ($\alpha$, $\delta$)       & ($\alpha$, $\delta$)     & ($\Delta \alpha$, $\Delta \delta$) \\
\hline
2007/324 &  \ \ (2.77, -10.80)  &  ---                   &  ---                  \\
2007/361 &  \ \ (2.35, -16.20)  &  \ (2.07,   -16.15)  & \ (0.28,  -0.05)    \\
2008/035 &  \ \ (3.85, -18.28)  &  \ (3.57,   -18.42)  & \ (0.28, \ 0.14)    \\
2008/063 &  \ \ (5.40, -20.35)  &  \ (5.26,   -20.50)  & \ (0.14, \ 0.15)    \\
2008/106 &  \  (10.07, -22.86)  &  ---                   &  ---                  \\
\hline
\multicolumn{4}{l}{$a$: Phase calibrations were done for L1448C and the results were } \\
\multicolumn{4}{l}{ transferred to J0319+3101. }\\
\multicolumn{4}{l}{$b$: Phase calibrations were done for J0319+3101 and the results were }\\
\multicolumn{4}{l}{ transferred to L1448C.} \\
\multicolumn{4}{l}{Note --- The position offsets are measured with respect to the } \\
\multicolumn{4}{l}{reference position of the masers in unit of milliarcseconds (mas).} \\
\end{tabular}
\end{center}
\end{table*}

\subsection{Astrometry of the H$_{2}$O masers in L1448C}

We have successfully determined the absolute positions of the maser spots 
by referring to the position reference source J0319+3101. 
In the data analysis, we performed phase-calibration on either 
the intense maser spot in L1448C or continuum source J0319+3101 as 
mentioned above. 
Table \ref{tab-prcompare1} compares the results of both methods. 
Except for the epoch 2007/324 and 2008/106, where L1448C could not be 
detected by phase-referencing to the J0319+3101, derived position offsets 
are consistent with each other. The uncertainties of about 0.3~mas 
(as summarized in the fourth column in Table \ref{tab-prcompare1}) would be 
a result of low signal-to-noise ratio of J0319+3101 or structure in the maser spots. 
The peak fluxes of J0319+3101 obtained by phase-referencing to 
the maser spots of L1448C recover 29\% and 52-55\% of those of 
self-calibrated images for 2007/324 and rest of 4 epochs, respectively. 
On the other hand, phase-referenced image of the maser spots only recover 
23-44\% of those of self-calibrated image for the epoch 2007/361, 2008/035, and 
2008/063. This is due to larger coherence loss 
in phase calibration on J0319+3101 with the lower signal-to-noise ratio. 
Thus, we hereafter employ the astrometric results obtained by phase-referencing 
to the reference maser spot in L1448C. 

\begin{figure*}[htb]
  \begin{center}
    \FigureFile(180mm,180mm){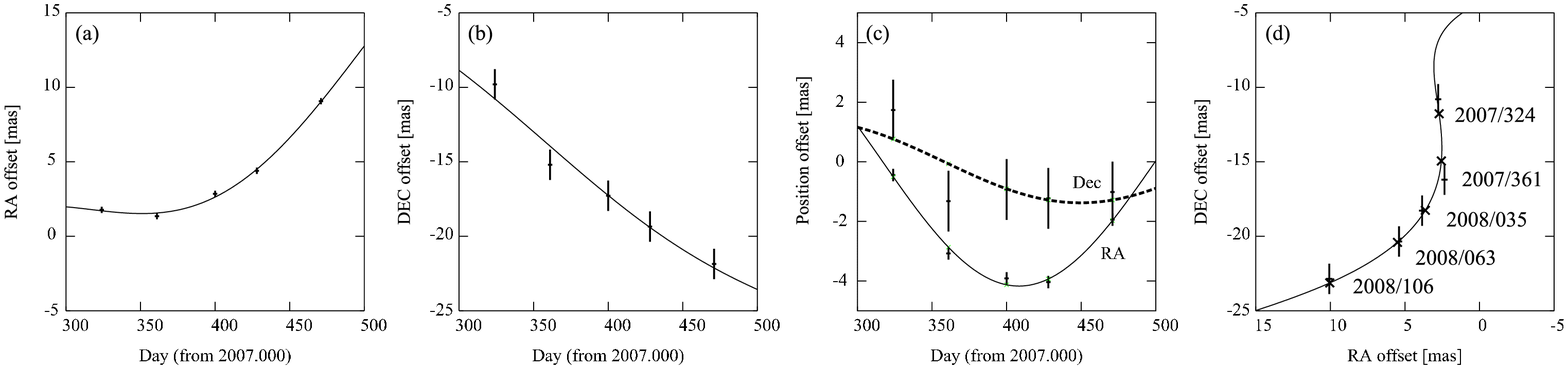}
  \caption{Position measurements of the maser spots at $v_{lsr}=20.6$~km~s$^{-1}$. 
   (a) The movement in right ascension as a function of time. 
   (b) The same as (a) but in declination. 
   (c) The same as (a) and (b) but the best fit proper motions are removed. 
   (d) The movement of the maser spots on the sky plane. 
   Solid lines represent the best fit model of the annual parallax. 
   The associated error bars, 0.21~mas and 1.02~mas 
   in right ascension and declination, respectively, are also plotted. 
   Small crosses in panel (c) indicate the expected position for each epoch 
   as labeled in the figure. }
  \label{fig-parallax}
  \end{center}
\end{figure*}

Figure \ref{fig-parallax} shows the position offsets of the masers 
as a function of observing epoch. 
The movement of the masers significantly deviates from a simple linear motion, 
and it can be fitted as a combination of linear proper motion and 
the annual parallax by the least-squares analysis. 
The results of the fitting are summarized in Table \ref{tab-results}. 
Using both the right ascension and declination data, 
the annual parallax of L1448C is derived to be 4.31$\pm$0.33~mas, 
corresponding to the distance of 232$\pm$18~pc from the Sun. 
The standard deviations of the post-fit residuals 
are $\sigma_{\alpha}$=0.19~mas and $\sigma_{\delta}$=0.93~mas 
in right ascension and declination, respectively. 

\begin{table}[tb]
\begin{center}
\caption{Results of the least-squares analysis for the annual parallax and proper motion measurements}
\label{tab-results}
\begin{tabular}{lc}
 \hline\hline
Parameter                         & Best fit value               \\ 
\hline
$\pi$ (mas)                                          &  4.31(33)      \\
$D$ (pc)                                              &   232(18)     \\
$\mu_{\alpha} \cos \delta$ (mas~yr$^{-1}$)  &  21.9(7)      \\
$\mu_{\delta}$ (mas~yr$^{-1}$)                  & -23.1(33)    \\
$\mu$ (mas~yr$^{-1}$)                            &  31.8(34)     \\
$v_{t}$ (km~s$^{-1}$)                               &  35.3(37)    \\
PA (degrees)                                         &  137          \\
$\sigma_{\alpha}$ (mas)                           & 0.19           \\
$\sigma_{\delta}$ (mas)                            &  0.93          \\
\hline \\
\multicolumn{2}{l}{Note --- Numbers in parenthesis represent} \\
\multicolumn{2}{l}{the errors in unit of the last significant digits. }\\
\multicolumn{2}{l}{$\sigma_{\alpha}$ and $\sigma_{\delta}$ are the 
the post-fit residuals. }\\
\end{tabular}
\end{center}
\end{table}

The larger residual in declination is due to position errors in 
the observations at 2007/361 and possibly 2007/324 as can be seen in 
Figures \ref{fig-parallax}(b) and (c). 
These errors are significantly larger than the formal errors in the Gaussian 
fitting of the maser spots, 0.03-0.1~mas. 
Therefore, we introduced the error floor of 0.21~mas and 1.02~mas in right 
ascension and declination, respectively, for all the results of the position 
measurements to make reduced $\chi^{2}$ to be unity in the least-squares 
analysis. These error floors represent the positional uncertainties in the 
present astrometric observations. 
The possible origin of these uncertainties is most likely due to the difference in 
the optical path lengths between the target and reference 
sources caused by the atmospheric zenith delay residual 
and/or a variability of the structure of the maser feature 
(see detailed discussions in \cite{honma2007}, \cite{hirota2007}, 
\cite{hirota2008a}, \cite{hirota2008b}).  

Based on the simulation by \citet{honma2008b}, positional errors due to 
the typical zenith delay residual in the VERA observations, 2~cm, 
can be estimated to be 0.03~mas and 0.01~mas in right ascension and declination, 
respectively, in the condition close to the present observations 
(i.e. the separation angle of 1.37~degrees, position angle on the sky from 
the target sources to the reference sources of 90~degrees, 
and the source declination of 15~degrees are employed). 
Although the assumed value of declination is different from that of L1448C, 
the estimated position offsets are much smaller than the observed results. 
Therefore, only the zenith delay residual would not be the cause of 
the large positional uncertainties unless unexpectedly large zenith delay residuals 
remain in the data. 

The larger positional uncertainties found in the H$_{2}$O masers associated 
with nearby low-mass YSOs 
\citep{hirota2007, imai2007, hirota2008a, hirota2008b} than in the distant 
massive YSOs (e.g. \cite{honma2007}) are possibly attributed to the source structure. 
In fact, such a variation of the internal maser structures have been observed 
directly by the recent VLBI observations of the H$_{2}$O masers in a nearby low-mass 
YSO, NGC~1333~IRAS4 \citep{marvel2008, desmurs2009}, with typical timescale of 
as short as 2-4~weeks. This may strongly affect the results of identifications 
for H$_{2}$O maser spots between two consecutive epochs separated by only 
1 month. 

Nevertheless, we can obtain the consistent result, $\pi$=4.30$\pm$0.37~mas 
and $\sigma_{\alpha}$=0.23~mas, 
even if only the right ascension data are used in the least-squares analysis. 
When we fit the data excluding those of the second epoch 2007/361, 
the resultant parallax value is consistent within the mutual error (4.23$\pm$0.29~mas). 
In this case, the post-fit residuals, $\sigma_{\alpha}$=0.15~mas and $\sigma_{\delta}$=0.39~mas 
in right ascension and declination, respectively, are reduced. 
Unfortunately, the masers associated with L1448C disappeared after the 6-month 
monitoring observations with VERA, which is also similar to the case 
for NGC~1333~SVS13A \citep{hirota2008a}. 
If the above error sources affect the position measurements as a random noise, 
the accuracy of the parallax measurement could be improved by increasing 
the number of observed epochs as well as the detected maser spots. 
Thus, we need to continue longer monitoring observations of the H$_{2}$O maser sources 
to improve the accuracy of the annual parallax measurements with VERA. 

\section{Discussions}

\subsection{Collimated H$_{2}$O maser jet from L1448C}

Both the spectra and distribution of the maser spots as shown in Figures 
\ref{fig-spectra} and \ref{fig-allspot} imply that the masers trace the base 
of the protostellar jet driven by the Class 0 source L1448C. 
In this section, we will discuss about the velocity structure of the 
H$_{2}$O masers. 

In Table \ref{tab-results}, we derived the absolute proper motion 
of the reference maser spot at the LSR velocity of 20.6~km~s$^{-1}$. 
In addition, we also derived the relative proper motions for other maser spots 
with respect to this reference spot. As a result, 
the northern spots at the LSR velocity of 7.1-7.7~km~s$^{-1}$ 
are found to be moving away from the reference spot. 
The average of the proper motions is 10.7~mas~yr$^{-1}$ or 11.9~km~s$^{-1}$ 
with the position angle of -34~degrees 
($\mu_{\alpha} \cos \delta$=-5.9~mas~yr$^{-1}$, $\mu_{\delta}$=8.9~mas~yr$^{-1}$) 
with respect to the reference spot. 
Based on the fact that the observed red-shifted jet lobe should be expanding 
toward the southeast direction \citep{girart2001}, the above motion could be 
naturally interpreted that the southern components are systematically moving 
toward the southeast direction with respect to the northern one. 
It is consistent with the expected position of the protostar located at the 
northwest of the masers as mentioned in the previous section. 
If this is the case, the maser spots with larger proper motion tends to be 
distributed away from the driving source. 

The radial velocity distribution of the H$_{2}$O masers shown in 
Figure \ref{fig-allspot} also suggests that the higher velocity components 
(20~km~s$^{-1}$) are distributed at the more distant positions from the 
powering source, which is possibly at the northwest of the masers, 
than the lower velocity components (7~km~s$^{-1}$). 
In addition, the LSR velocities within the southern maser feature gradually 
increase from 19~km~s$^{-1}$ at the northern end to 21~km~s$^{-1}$ 
at the southern end (Figure \ref{fig-allspot}(c)). 
Therefore, both the proper motion and radial velocity structure might exhibit 
an apparent acceleration of the jet driven by the protostar L1448C. 

Based on the velocity shift in both the proper motion and the 
radial velocity between the northern and southern features, 
we roughly estimate the inclination angle of the maser jet to 
the line of sight to be 43~degrees. 
This is smaller than that of the kinematic model proposed 
for large-scale bipolar outflow, 70~degrees, 
even if the opening angle of the red-shifted jet lobe, 30~degrees, is taken 
into account \citep{bachiller1995, girart2001}. 
Although our estimation of the inclination angle 
would contain large uncertainty, this result suggests an evidence 
for the precession of the jet axis. 

As shown in Figure \ref{fig-allspot}, the maser spots are aligned along 
the northwest-southeast direction. We obtained the position angle of 
the alignment of the H$_{2}$O maser features to be -15~degrees on average. 
This is in good agreement with that of the large-scale collimated 
molecular outflow from L1448C, -21~degrees (e.g. \cite{bachiller1995}), 
while slightly different from the direction of the proper motion as mentioned 
above, -34~degrees. 
According to the Spitzer IRAC2 image in Figure \ref{fig-spitzer}, 
the infrared jet is almost parallel to the H$_{2}$O masers while it 
show slightly curved structure at the northern part of the jet. 
This is thought to be an evidence for the interaction with the jet and ambient gas 
around another protostar L1448N \citep{bachiller1995}. 
In addition, \citet{hirano2010} found the deflection of the both 
blue-shifted and red-shifted SiO jet 
with almost point-symmetry with respect to its driving source, which is 
possibly caused by the orbital motion of the binary system. 
Thus, it is likely that the variation in the position angles of the jets 
derived from the different tracers would be due to the precession of the jet axis. 
It should be noted that a possible evidence for the precession is also 
reported for the H$_{2}$O maser jet in NGC~1333~IRAS4 
\citep{marvel2008, desmurs2009}. 

We could evaluate the timescale of the maser jet 
to be at least 9~years based on the velocity difference 
in the proper motion between the northern and southern maser features. 
If the maser spots are accelerated at a constant rate 
(10.7~mas~yr$^{-1}$/9~yr=1.2~mas~yr$^{-1}$~yr$^{-1}$), the proper motion 
would increase by 0.6~mas~yr$^{-1}$ within the monitoring period of 0.5~year. 
Although such acceleration might affect the accuracy of 
the annual parallax and proper motion measurements, 
we could not see significant deviation from the linear proper motion 
with respect to the reference spot within the fitting errors. 
It would mean that the physical gas clumps appeared as the H$_{2}$O maser 
features are not really accelerated for 9~years but they just represent 
the velocity structure of the jet showing apparent acceleration. 

The kinematics of the H$_{2}$O maser jet and its time scale are still 
uncertain mainly because we could not detect the blue-shifted H$_{2}$O 
maser features. 
In order to reveal the launching mechanism of the protostellar jet 
associated with L1448C, which might be the binary system 
\citep{hirano2010}, proper motion measurements of both red- and 
blue-shifted masers would be a key issue. 

\subsection{Overall structure of the Perseus molecular cloud complex}

The present astrometric observations with VERA imply that 
the distance to L1448C is most likely 232~pc, which is in 
excellent agreement with that of our previous results for NGC~1333~SVS13A, 
235$\pm$18~pc \citep{hirota2008a}. 
Because the angular separation between these two sources, 1~degree, 
corresponds to the linear size of 5~pc, the similar extent along the 
line of sight is also plausible. 
Although we could not distinguish the difference in their distances, 
more precise parallax measurements will allow us to reveal 
the depth of the molecular cloud. 

Our distance measurements are consistent with the photometric 
distance to NGC~1333 reported by \citet{cernis1990} of 220~pc with an 
uncertainty of 25\% rather than the larger value of about 300~pc 
(e.g. \cite{herbig1983}; \cite{dezeeuw1999}). 
Therefore, the present result provides a constraint on the distance to 
the western part of the Perseus molecular cloud complex to be closer value. 

\begin{figure}[htb]
  \begin{center}
    \FigureFile(85mm,85){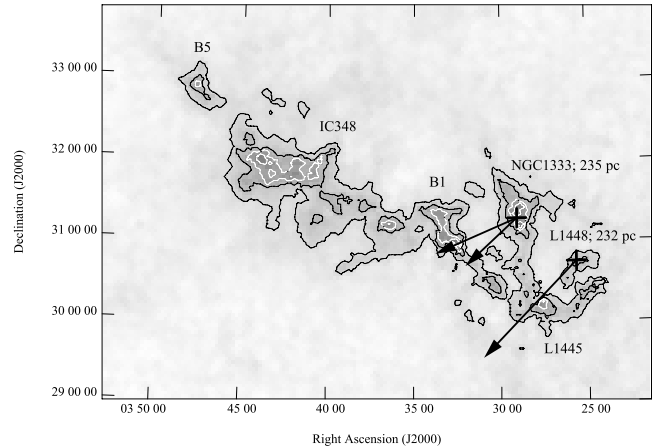}
  \caption{Absolute proper motions of the H$_{2}$O masers in L1448C and NGC~1333~SVS13A
   \citep{hirota2008a}. Grey scale shows the visual extinction ($A_{V}$) map 
    derived from the 2MASS data \citep{ridge2006}. Contour levels are $A_{V}$=3, 5, 7, and 9 magnitude. 
   Crosses and arrows represent the position of the maser sources and 
   their absolute proper motion vectors, respectively.  }
  \label{fig-av}
  \end{center}
\end{figure}

\citet{enoch2006} suggested that a single distance for the whole of 
the Perseus molecular cloud complex might not be appropriate 
although they adopted the distance of 250~pc for the entire area of 
the Perseus region. 
According to the photometric observations by \citet{cernis1990} and 
\citet{cernis1993}, there exists a gradient in the distances across the 
Perseus molecular cloud complex. NGC 1333 is proposed to be 
the nearest cloud at a distance of 220~pc \citep{cernis1990} while 
IC348 is more distant, 300~pc \citep{cernis1993}. 
The latter value is consistent with that of the Hipparcos result 
(318$\pm$27~pc; \cite{dezeeuw1999}). 
Thus, it would be interesting to carry out VLBI astrometry of the H$_{2}$O maser 
sources and radio emitting T-Tauri stars (e.g. \cite{loinard2005}) 
in other Perseus clouds, in particular for 
the eastern part of the complex such as B1, IC348, and B5 (see Figure \ref{fig-av}). 

As listed in Table \ref{tab-results}, 
the absolute proper motion of the reference maser spot is 
31.8~mas~yr$^{-1}$ or 35.3~km~s$^{-1}$ toward southeast with the position angle 
of 137 degrees. The direction of the proper motion seems to be 
consistent with those of NGC1333~SVS13A \citep{hirota2008a} 
as plotted in Figure \ref{fig-av}. It is also reported that the average 
proper motion for the radio continuum sources in NGC~1333 agrees well 
with those of the H$_{2}$O masers \citep{carrasco2008}. 
These systematic motion could be indicative of the proper motion of 
the Perseus molecular cloud itself. 
The larger proper motions for L1448C than NGC1333~SVS13A might be 
caused by the contamination of the jet motion. 
High-resolution interferometric observations of continuum emission 
from YSOs at the centimeter, millimeter, and 
submillimeter wavelengths with EVLA and ALMA will be crucial to distinguish 
the contribution from proper motions of outflows, stars, and host clouds. 

\vspace{12pt}
We are grateful to the staff of all the VERA stations 
for their assistance in observations. 
TH is financially supported by Grant-in-Aids from 
the Ministry of Education, Culture, Sports, Science and Technology 
(13640242, 16540224, and 20740112).

\end{document}